\documentclass[submission,copyright,creativecommons]{eptcs}
\providecommand{\event}{ICLP \& LPNMR Doctoral Consortium 2024} 

\usepackage{iftex}

\ifpdf
  \usepackage{underscore}         
  \usepackage[T1]{fontenc}        
\else
  \usepackage{breakurl}           
\fi

\title{A Category-Theoretic Perspective on Approximation Fixpoint Theory\thanks{This PhD project is supported by Fonds Wetenschappelijk Onderzoek -- Vlaanderen (project G0B2221N), with supervisors Bart Bogaerts and Marc Denecker.} }
\author{Samuele Pollaci
\institute{Vrije Universiteit Brussel \\ Brussels, Belgium}
\institute{Katholieke Universiteit Leuven \\
Leuven, Belgium}
\email{Samuele.Pollaci@vub.be}
}
\def\titlerunning{A Category-Theoretic Perspective on Approximation Fixpoint Theory }
\def\authorrunning{S. Pollaci}
\begin{document}
\maketitle

\begin{abstract}
Approximation Fixpoint Theory (AFT) was founded in the early 2000s by Denecker, Marek, and Truszczyński as an abstract algebraic framework to study the semantics of non-monotonic logics. 
Since its early successes, the potential of AFT as a unifying semantic framework has become widely recognised, and the interest in AFT has gradually increased, with applications now ranging from foundations of database theory to abstract argumentation.  The non-monotonic constructive processes that occur in many more areas of computer science, together with their associated semantic structures, can be successfully studied using AFT, which greatly simplifies their characterizations.
The goal of my research is to take a step towards the lifting of AFT into a more general framework for constructive knowledge. 

\end{abstract}

\section{Introduction and Related Work}

Approximation Fixpoint Theory (AFT) was founded in the 2000s by Denecker, Marek, and Truszczyński \cite{DMT00ApproximationsStableOperatorsWell-FoundedFixpointsApplications} as an extension of Tarski’s fixpoint theory  to study the semantics of  non-monotonic logics, like default logic (DL), autoepistemic logic (AEL) and logic programming (LP). In recent years, interest in AFT has gradually increased, with applications now ranging from foundations of database theory to abstract argumentation. Motivated by the success of AFT in this wide range of applications, this project aims at laying the foundation for an extention of AFT from a useful tool in the area non-monotonic logics into a general algebraic theory of constructive knowledge. 

In the 1980s and 90s, the area of non-monotonic reasoning (NMR) saw fierce debates about formal semantics. In the subareas of DL, AEL and LP, researchers sought to formalize common-sense intuitions about knowledge of introspective agents. The main contribution of AFT was to demonstrate that, by moving to an algebraic setting, the common principles behind the concepts in these languages can be isolated and studied in a general way. This breakthrough allowed results that were achieved in the context of one of these languages to be easily transferred to another \cite{DBLP:journals/amai/Truszczynski06,VWMD07PredicateIntroductionLogicsFixpointSemanticsII}. 

The core ideas of AFT are relatively simple: we are interested in fixpoints of an operator $O$ on a given lattice $\langle L,\leq\rangle$. 
For monotonic operators, Tarski's theory guarantees the existence of a least fixpoint, which is of interest in many applications. 
For non-monotonic operators, the existence of fixpoints is not guaranteed; and even if fixpoints exist, it is not clear which would be ``good'' fixpoints.
AFT generalizes Tarki's theory for monotonic operators by making use of a so-called \emph{approximating operator}; this is an operator $A: L^2\to L^2$, that operates on $L^2$, and that is monotonic with respect to the precision order $\leq_p$ (defined by $(x,y)\leq_p(u,v)$ if $x\leq u$ and $v \leq y)$). The intuition is that elements of $L^2$ approximate elements of $L$:  $(x,y)\in L^2$ approximates $z$ if $x\leq z\leq y$, i.e.\ when $x\leq y$, the tuple $(x,y)$ can be thought of as an interval in $L$. 
Given such an approximator, AFT defines several types of fixpoints (supported fixpoints, a Kripke-Kleene fixpoint, stable fixpoints, a well-founded fixpoint) of interest. 
In several fields of non-monotonic reasoning, 
it is relatively straightforward to define an approximating operator 
and
it turns out that the different types of fixpoints then correspond to existing semantics. 
In this way, AFT clarifies on the one hand how different semantics in a single domain relate, and on the other hand what the relation is between different (non-monotonic) logics. 

%

Since its early successes, the potential of AFT as a unifying semantic framework has become widely recognised. It has been applied to a multi-agent extension of AEL \cite{VCBD16DistributedAutoepistemicLogicApplicationAccessControl}, to Dung’s abstract argumentation frameworks (AFs), and abstract dialectical frameworks (ADFs) \cite{S13Approximatingoperatorssemanticsabstractdialecticalframeworks}. It has also been applied to extensions of LP with aggregates \cite{DPB01UltimateWell-FoundedStableSemanticsLogicPrograms}, with HEX-atoms \cite{AEF13HexSemanticsApproximationFixpointTheory}, with higher-order functions \cite{CRS18ApproximationFixpointTheoryWell-FoundedSemanticsHigher-Order}, and with description logics \cite{LBCYF16FlexibleApproximatorsApproximatingFixpointTheory}. It served as the foundation of the causal logic FO(C) \cite{BVDV14FOCKnowledgeRepresentationLanguageCausality}  and was used to characterize active integrity constraints \cite{BC18Fixpointsemanticsactiveintegrityconstraints}. In all these domains, the power of AFT allows a variety of complex semantic structures to be characterized by surprisingly simple approximating operators. In comparison to a direct definition of the semantic structures, the AFT approach therefore greatly reduces the risk of error, while significantly simplifying the mathematical study of these structures. Interest in these application areas has also driven the theoretical development of AFT in new directions \cite{LBCYF16FlexibleApproximatorsApproximatingFixpointTheory, CRS18ApproximationFixpointTheoryWell-FoundedSemanticsHigher-Order,BVD15Groundedfixpointstheirapplicationsknowledgerepresentation,BVD18Safeinductionstheirapplicationsknowledgerepresentation,DV07Well-FoundedSemanticsAlgebraicTheoryNon-monotoneInductive}. 

The recent broad interest in AFT and the resulting research output provide a unique opportunity to put AFT on the map as a general algebraic theory of constructive knowledge. Throughout computer science, Tarski’s fixpoint theory is used because it allows a potentially complex object (the least fixpoint of a monotone operator) to be constructed as the limit of a simple iteration process. In many applications, however, this simple monotone construction process does not suffice. In such cases, it is sometimes possible to devise other, derived operators that do not operate on the basic semantic space but on a more complex space. The core of AFT lies in the study of the general principles that underlie such approximations. Such non-monotone construction processes occur in many more areas of computer science, including formal verification, functional programming, and database theory. The goal of this project is to develop a single unified framework to study them, thereby bringing two important benefits. 
First, a general theory provides confidence in the correct characterization of the processes and resulting semantics. Non-monotone construction processes can be highly complex and developing a new formalization of such a process from scratch is a difficult, time-consuming and error-prone task. This advantage of AFT has already been convincingly demonstrated for logic programs with aggregates, where direct definitions of the desired semantic structures are highly complex, while the use of AFT requires nothing more than the definition of a (three-valued) truth evaluation function for the aggregates. 
Second, once the constructive process has been correctly characterized, AFT offers a powerful set of algebraic tools to analyze the process and its limit. For instance, a general toolset for the study of modularity properties was developed, generalizing several known concepts from AEL, DL and LP \cite{T06Stronguniformequivalencenonmonotonictheories-,VGD06SplittingoperatorAlgebraicmodularityresultslogics,VWMD07PredicateIntroductionLogicsFixpointSemanticsI,VWMD07PredicateIntroductionLogicsFixpointSemanticsII}. As a second example, Truszczyński \cite{T06Stronguniformequivalencenonmonotonictheories-} developed tools for the study of strong equivalence, which immediately transfer to other fields, where such notions are now studied independently  \cite{OW11Characterizingstrongequivalenceargumentationframeworks}. 
AFT greatly simplifies the characterization and subsequent study of constructive processes and their associated semantic structures. By lifting AFT into a more general framework for constructive knowledge, we bring these benefits to a wide range of application areas in computer science.

\section{Scientific Research Goals}

My PhD research activity fits into the framework of the AFTACK project (project G0B2221N) supported by Fonds Wetenschappelijk Onderzoek -- Vlaanderen. The global goal of this project is to take the next big leap forward for AFT by lifting the results obtained for various specific application domains into a general framework for constructive knowledge. To achieve this, three major types of advancements are needed.
\begin{enumerate}
\item[A.]
General approximation spaces: A construction process consists of a series of approximative objects. In AFT, these are elements $(x, y)$ of the bilattice $L^2$, which correspond to intervals $[x,y]= \{z \in L \mid x \leq z \leq y\}$ in the original lattice. In various scenarios, intervals are not refined enough and a more general approximation space is needed.
\item[B.]
General processes: AFT was built for processes typically found in non monotonic reasoning. When moving beyond this scope, the basic concepts behind constructive processes remain the same, but certain key differences nevertheless arise. For instance, while processes in AFT typically construct relations over a given set, domain theory \cite{S72Continuouslattices} considers processes that simultaneously construct the relations and the set over which they are defined. 
\item[C.]
Analysis of processes: The success of AFT is for a large part due to the rich algebraic toolkit it offers to analyze the defined processes and semantics. In parallel with extending the range of spaces and processes, this toolkit needs to be extended as well. 
\end{enumerate}

These three goals are materialized in five concrete research objectives. In my research, I am going to focus primarily on three of them:
\begin{enumerate}
	\item Approximation Spaces (A). Develop a generalization of AFT where the approximations can be complex mathematical structures, instead of simple intervals of lattice elements. In several applications, the limitation to intervals was recognized as a key limiting factor \cite{CRS18ApproximationFixpointTheoryWell-FoundedSemanticsHigher-Order,LBCYF16FlexibleApproximatorsApproximatingFixpointTheory,BVD16Well-FoundedSet-InductionsLocallyMonotoneOperators,B19WeightedAbstractDialecticalFrameworksthroughLens}. The generalisation should be general enough to cover these domains. 
	\item Recursively defined domains and higher-order functions (A,B). Develop extensions of AFT and domain theory suitable to define recursive higher-order functions and predicates. Preliminary experiments have shown that AFT is unsuitable for recursive definitions of higher-order functions \cite{DvBJD16CompositionalTypedHigher-OrderLogicDefinitions}. For monotonic recursively defined functions, a solution is provided in domain theory. The objective is to extend domain theory \cite{S72Continuouslattices} with the fixpoint notions of AFT to handle non-monotonically defined recursive functions and predicates. 
	\item Explanations for AFT (C). In many domains, it is not only important to reach the right conclusions, but also to explain why they are correct. For instance, in causality, this is the question of actual causation \cite{H16AppropriateCausalModelsstabilityCausation}. In the context of constructive knowledge, this question can be posed as “Why does a property hold in the constructed object?”. Achieving explainability is especially important in the light of the EU General Data Protection Regulation, article 22 of which requires that all AI with an impact on human lives needs to be accountable. Therefore, we want to obtain a principled approach towards explanations in AFT, which will immediately be applicable to all logics captured by AFT.  
\end{enumerate}

The impact of this project is spread over many different domains. In fact, within each new application domain, valuable lessons can be learnt from AFT, as has been witnessed before, for instance in the context of default logic, weighted argumentation, and active integrity constraints. The largest short-term impact is probably found in Objective 2 (Recursively defined domains and higher-order functions). If successful, this project will on the one hand be a bridge between functional programming and non-monotonic reasoning, and on the other hand, will provide the semantic foundations for a new class of function definitions by lifting the restriction that function definitions need to be monotonic in the definedness order. Finally, explainability is an important topic in many of the application domains of AFT right now. By studying this once, algebraically, we lay the foundations for work on explanations in each of these research areas.

\section{Research Methodology}

In the following, I break down the three objective presented above into smaller work packages (WP). This subdivision follows the one proposed for the AFTACK project. In particular, WP2 will be the main core of my research activity. At the same time, I will collaborate with other members of the research group on (parts of) the other work packages. More details are provided in Section \ref{sec:current}.

\subsection{WP1 Approximation Spaces} 

To develop a notion of an approximation space of a given lattice $\langle L, \leq\rangle$, we need a mathematical structure equipped with a truth order $\leq$ and a precision order $\leq_p$, on which we define a generalization of an approximating operator. It is important that the space is equipped with enough structure to allow the construction of key concepts of AFT, such as the stable operator, whose fixpoints determine the partial stable model semantics and the well-founded semantics \cite{DMT00ApproximationsStableOperatorsWell-FoundedFixpointsApplications}. We first tackle this problem for two specific structures. We expect the lessons learned there to prove valuable when finally developing approximation spaces in WP1.3. 

\begin{itemize}
\item[WP1.1.] We first consider the use of the powerset of $L$ as an approximations space. Based on a preliminary analysis, we conjecture that it is impossible to define a suitable truth order on the set of all subsets $S$ of $L$. Therefore, we focus on sets “without holes”, formally, sets S such that whenever $x,z \in S$ and $x < y < z$, also $y \in S$. The most important challenge is defining a stable operator in this context, and then studying the resulting semantics. 
\item[WP1.2.] Weighted Abstract Dialectical Frameworks (wADFs) \cite{BSWW18WeightedAbstractDialecticalFrameworks} were originally defined in the context of an arbitrary set with a precision order (a complete partial order) but no truth order. In this work package, we research (i) how to add a truth order to wADFs, and (ii) which restrictions on the respective orders we need in order to generalize stable and well-founded semantics to wADFs. 
\item[WP1.3.] Now, we are concerned with formally defining the notion of an approximation space. This should be a mathematical structure $\langle L,\leq ,\leq_p\rangle$ of approximations of a lattice $\langle L, \leq $ such that (i) there is a $\leq$-preserving injection from $L$ into $L$; (ii) we can generalize the definition of an approximating operator $A \colon L^2 \to L^2$ on the bilattice $L^2$ to approximating operators $A \colon L \to L$ on the approximation space $L$; (iii) we can generalize the key concepts of AFT (such as the stable operator) to $L$; and (iv) the generalizations cover at least the approximations from WP1.1, and WP1.2, and possibly other extensions of AFT \cite{CRS18ApproximationFixpointTheoryWell-FoundedSemanticsHigher-Order,LBCYF16FlexibleApproximatorsApproximatingFixpointTheory}. 
\item[WP1.4.] We investigate what it means for an operator on an approximation space to be stratified or modular. We also investigate which types of fixpoints of the generalized AFT behave well under such stratification \cite{VGD06SplittingoperatorAlgebraicmodularityresultslogics}.
\end{itemize}

\subsection{WP2 Recursively defined domains and higher-order functions} 

Defining semantics for definitions of higher-order objects is challenging for typed languages and even more so for untyped ones. Semantics for typed and untyped monotone higher-order function definitions are developed in denotational semantics for lambda calculus and functional programming languages \cite{T91Semanticsprogramminglanguages}; one approach is domain theory \cite{S72Continuouslattices}. Domain theory  was the first approach to provide semantics for the untyped lambda calculus and allows for more complex semantic spaces than AFT, with rich classes of approximate objects, suitable for inductive and coinductive constructions of higher-order objects and infinite data structures. It simultaneously constructs approximations of these complex objects and of the functions operating on them, resulting in a space of self-applicable functions including a fixpoint operator through which recursive definitions are given meaning. In another sense, domain theory is more limited than AFT: it only considers continuous (hence, monotone) operators and definitions. Here, we combine the strengths of both approaches. 

\begin{itemize}
	\item[WP2.1.] We study approximation spaces in a category-theoretic setting. In particular, we investigate which classes of approximation spaces form a Cartesian-closed category. One reason we are interested in this is that a Cartesian-closed category provides a foundation for applying AFT to higher-order definitions, as it provides a systematic construction of approximation domains for higher-order concepts from those of base concepts. We undertake this investigation for different notions of approximations.  
	\item[WP2.2.] Here, we exploit the framework of WP2.1 to develop languages and semantics of non-monotone definitions of higher-order sets and functions. The framework supports this (i) by providing construction of approximation domains for higher-order concepts; (ii) by imposing abstract conditions on language constructs that ensure that they are well-behaved (being morphisms in the category); (iii) by being generic in the underlying order, allowing to combine different orders (e.g., simultaneous definition of sets and functions).The range of notions of approximation space in the framework should offer an enriched spectrum with different trade-offs between precision, mathematical complexity, and computational complexity. 
	\item[WP2.3.] In AFT, the approximator is strictly higher in the set-theoretical hierarchy than the objects on which it operates. A powerful property of domain theory is that this is not the case: the objects it constructs can be applied as functions on all constructed objects, including itself. This property is key to defining semantics of untyped lambda calculus. Dana Scott’s construction of such a domain is one of the great achievements in the theory of programming languages, but depends on the $\leq_d$-continuity (and hence, monotonicity) of operators of definitions, a condition we seek to relinquish. In this final work package, we examine whether it is possible to partially lift the limitation of $\leq_d$-monotonicity of domain theory and, at the same time, lift AFT’s limitation that prevents its use for self-application. 
\end{itemize}

\subsection{WP3 Explanations for AFT} 

To bring explainability to AFT, we start from the theory of justifications \cite{DBS15FormalTheoryJustifications}. Like AFT, this is a unifying theory that can characterize semantics of various formalisms. However, it is more specific, in the sense that it does not consider operators on an arbitrary lattice, but is focused on the special case of a powerset lattice. This allows justification theory to build detailed explanations of why each element of a set $X \in L$ belongs to $X$. Nevertheless, there are strong correspondences between AFT and justification theory, including similar notions of duality. The exact relationship remains largely unexplored. 

\begin{itemize}
	\item[WP3.1.]  Instead of starting from an operator, justification theory starts from a ``justification frame", from which an operator can be derived. These frames can be nested, resulting in “nested justification systems”. In this work package, we research the relationship between the concepts of non-nested justification theory and those of AFT. In particular, we verify the conjecture that for all main “branch evaluations” (a concept from justification theory), the semantics induced by justification theory coincides with the equally named fixpoint(s) in AFT. In other words, this WP provides an embedding of non-nested justification theory into AFT. While this result is valuable in itself, the more interesting question is how to generalize AFT using notions from justification theory, which is what we do next. 
	\item[WP3.2.] In WP3.1, we consider only the operators on powerset lattices induced by justification theory. In this work package, we research (i) how to generalize the notion of justification to work for an approximator over an arbitrary complete lattice, or more general, approximation space, (ii) how to automatically obtain justifications from operators (or rather, from approximators), and (iii) how these obtained justifications translate back to justification theory. While the main focus of this work package is on extending approximation fixpoint theory, and bringing justifications to its various fields of application, the results achieved here also have a significant impact on the theory of justifications itself. Indeed, one consequence of these results would be that we now also obtain justifications for so-called ultimate semantics \cite{DMT04Ultimateapproximationapplicationnonmonotonicknowledgerepresentation}.  
	\item[WP3.3.] As mentioned, justification theory allows justification frames to be nested. This results in an elegant way of capturing the semantics of, for instance, nested least and greatest fixpoint definitions. Now, we research how to achieve the same in AFT. In other words, we develop a suitable notion of nested approximations, where a single “step” of a high-level approximation may include an entire fixpoint construction of a lower-level approximation. In addition, it should be possible to choose, for each of the different levels independently which are the fixpoints of interest. 
	\item[WP3.4.] In the context of justification theory, we recently defined notions of duality, one of which is induced by inverting the truth order. In general, we expect that such notions also show up in AFT and that many properties have interesting dual variants, for instance so-called symmetric approximators would be self-dual. In this work package we perform a complete study of duality in AFT in general: we identify dual properties and study which new results we obtain by exploiting them. 
\end{itemize}

The research goals of the AFTACK project in which my research activity fits are both fundamental and ambitious. It is therefore probable that we will not be able to meet every single one of them. However, this need not be a problem, since each of the individual steps towards the desired goals is in itself innovative and will lead to publications in high-impact conferences and journals. Moreover, the WPs are structured in such a way that there is little interdependence between them. 
The most central WP is that on approximation spaces (WP1). The research line on explanations (WP3) has a minor dependency on WP1, in the sense that it needs to be studied in the general context of approximation spaces in order to have maximum impact. However, even in the unlikely worst case that the WP on approximation spaces should fail completely, studying the topics in the context of the bilattice of standard AFT still provides novel and interesting results. The research line on recursive functions has the strongest dependency on approximation spaces, in the sense that we expect that the bilattice will not suffice for tackling this topic. Nevertheless, even if we cannot construct approximation spaces that are general enough to meet all of our stated goals for that WP, we can still develop a less ambitious extension of the bilattice that will be enough to allow the work on recursive functions to move forward.

\section{Current Status and Future Plan}\label{sec:current}

So far, I have been mainly working on WP2 for AFT for higher-order definitions, specifically WP2.1. and WP2.2.. In particular, a paper on the stable semantics for higher-order logic programming has been accepted for ICLP2024, and another paper on the generalization of approximation spaces for higher-order objects using the tools of Category Theory is under review for LPNMR2024. The latter paper is also contributing to WP1.1 and WP1.3 as it proposes a novel, general notion of possible approximation spaces.  Continuing this line of work, I am currently studying the relation between our new research outputs and previous approaches, like \cite{CRS18ApproximationFixpointTheoryWell-FoundedSemanticsHigher-Order}, both from a theoretic perspective and from a more computational one, regarding the complexity in finding the models of a logic program. As next step, I plan to work towards the completion of both WP1 and WP2, in particular by tackling WP1.2. on wADFs, WP1.4., and WP2.3..  The study of AFT for wADFs may lead the way for the application of AFT to bound-founded ASP \cite{ACS13Stablemodelsemanticsfoundedbounds}, assumption-based argumentation \cite{BDKT97AbstractArgumentation-TheoreticApproachDefaultReasoning}, weighted argumentation \cite{DHMPW11WeightedargumentsystemsBasicdefinitionsalgorithms}, probabilistic argumentation \cite{DT10TowardsProbabilisticArgumentationJury-basedDisputeResolution}, and social argumentation \cite{LM11SocialAbstractArgumentation}.

\nocite{*}
\bibliographystyle{eptcs}
\bibliography{dc}

\begin{thebibliography}{10}
\providecommand{\bibitemdeclare}[2]{}
\providecommand{\surnamestart}{}
\providecommand{\surnameend}{}
\providecommand{\urlprefix}{Available at }
\providecommand{\url}[1]{\texttt{#1}}
\providecommand{\href}[2]{\texttt{#2}}
\providecommand{\urlalt}[2]{\href{#1}{#2}}
\providecommand{\doi}[1]{doi:\urlalt{https://doi.org/#1}{#1}}
\providecommand{\eprint}[1]{arXiv:\urlalt{https://arxiv.org/abs/#1}{#1}}
\providecommand{\bibinfo}[2]{#2}

\bibitemdeclare{inproceedings}{AEF13HexSemanticsApproximationFixpointTheory}
\bibitem{AEF13HexSemanticsApproximationFixpointTheory}
\bibinfo{author}{Christian \surnamestart Antic\surnameend},
  \bibinfo{author}{Thomas \surnamestart Eiter\surnameend} \&
  \bibinfo{author}{Michael \surnamestart Fink\surnameend}
  (\bibinfo{year}{2013}): \emph{\bibinfo{title}{Hex Semantics via Approximation
  Fixpoint Theory}}.
\newblock In: {\slshape \bibinfo{booktitle}{Logic Programming and Nonmonotonic
  Reasoning, 12th International Conference, {LPNMR} 2013, Corunna, Spain,
  September 15-19, 2013. Proceedings}}, pp. \bibinfo{pages}{102--115},
  \doi{10.1007/978-3-642-40564-8\_11}.

\bibitemdeclare{article}{ACS13Stablemodelsemanticsfoundedbounds}
\bibitem{ACS13Stablemodelsemanticsfoundedbounds}
\bibinfo{author}{Rehan~Abdul \surnamestart Aziz\surnameend},
  \bibinfo{author}{Geoffrey \surnamestart Chu\surnameend} \&
  \bibinfo{author}{Peter~J. \surnamestart Stuckey\surnameend}
  (\bibinfo{year}{2013}): \emph{\bibinfo{title}{Stable model semantics for
  founded bounds}}.
\newblock {\slshape \bibinfo{journal}{Theory Pract. Log. Program.}}
  \bibinfo{volume}{13}(\bibinfo{number}{4-5}), pp. \bibinfo{pages}{517--532},
  \doi{10.1017/S147106841300032X}.

\bibitemdeclare{inproceedings}{B19WeightedAbstractDialecticalFrameworksthroughLens}
\bibitem{B19WeightedAbstractDialecticalFrameworksthroughLens}
\bibinfo{author}{Bart \surnamestart Bogaerts\surnameend}
  (\bibinfo{year}{2019}): \emph{\bibinfo{title}{Weighted Abstract Dialectical
  Frameworks through the Lens of Approximation Fixpoint Theory}}.
\newblock In: {\slshape \bibinfo{booktitle}{The Thirty-Third {AAAI} Conference
  on Artificial Intelligence, {AAAI} 2019, The Thirty-First Innovative
  Applications of Artificial Intelligence Conference, {IAAI} 2019, The Ninth
  {AAAI} Symposium on Educational Advances in Artificial Intelligence, {EAAI}
  2019, Honolulu, Hawaii, {USA}, January 27 - February 1, 2019}}, pp.
  \bibinfo{pages}{2686--2693}, \doi{10.1609/aaai.v33i01.33012686}.

\bibitemdeclare{article}{BC18Fixpointsemanticsactiveintegrityconstraints}
\bibitem{BC18Fixpointsemanticsactiveintegrityconstraints}
\bibinfo{author}{Bart \surnamestart Bogaerts\surnameend} \&
  \bibinfo{author}{Lu{\'{\i}}s \surnamestart Cruz{-}Filipe\surnameend}
  (\bibinfo{year}{2018}): \emph{\bibinfo{title}{Fixpoint semantics for active
  integrity constraints}}.
\newblock {\slshape \bibinfo{journal}{Artif. Intell.}} \bibinfo{volume}{255},
  pp. \bibinfo{pages}{43--70}, \doi{10.1016/j.artint.2017.11.003}.

\bibitemdeclare{article}{BVD15Groundedfixpointstheirapplicationsknowledgerepresentation}
\bibitem{BVD15Groundedfixpointstheirapplicationsknowledgerepresentation}
\bibinfo{author}{Bart \surnamestart Bogaerts\surnameend},
  \bibinfo{author}{Joost \surnamestart Vennekens\surnameend} \&
  \bibinfo{author}{Marc \surnamestart Denecker\surnameend}
  (\bibinfo{year}{2015}): \emph{\bibinfo{title}{Grounded fixpoints and their
  applications in knowledge representation}}.
\newblock {\slshape \bibinfo{journal}{Artif. Intell.}} \bibinfo{volume}{224},
  pp. \bibinfo{pages}{51--71}, \doi{10.1016/j.artint.2015.03.006}.

\bibitemdeclare{article}{BVD16Well-FoundedSet-InductionsLocallyMonotoneOperators}
\bibitem{BVD16Well-FoundedSet-InductionsLocallyMonotoneOperators}
\bibinfo{author}{Bart \surnamestart Bogaerts\surnameend},
  \bibinfo{author}{Joost \surnamestart Vennekens\surnameend} \&
  \bibinfo{author}{Marc \surnamestart Denecker\surnameend}
  (\bibinfo{year}{2016}): \emph{\bibinfo{title}{On Well-Founded Set-Inductions
  and Locally Monotone Operators}}.
\newblock {\slshape \bibinfo{journal}{{ACM} Trans. Comput. Log.}}
  \bibinfo{volume}{17}(\bibinfo{number}{4}), p.~\bibinfo{pages}{27},
  \doi{10.1145/2963096}.

\bibitemdeclare{article}{BVD18Safeinductionstheirapplicationsknowledgerepresentation}
\bibitem{BVD18Safeinductionstheirapplicationsknowledgerepresentation}
\bibinfo{author}{Bart \surnamestart Bogaerts\surnameend},
  \bibinfo{author}{Joost \surnamestart Vennekens\surnameend} \&
  \bibinfo{author}{Marc \surnamestart Denecker\surnameend}
  (\bibinfo{year}{2018}): \emph{\bibinfo{title}{Safe inductions and their
  applications in knowledge representation}}.
\newblock {\slshape \bibinfo{journal}{Artif. Intell.}} \bibinfo{volume}{259},
  pp. \bibinfo{pages}{167--185}, \doi{10.1016/j.artint.2018.03.008}.

\bibitemdeclare{article}{BVDV14FOCKnowledgeRepresentationLanguageCausality}
\bibitem{BVDV14FOCKnowledgeRepresentationLanguageCausality}
\bibinfo{author}{Bart \surnamestart Bogaerts\surnameend},
  \bibinfo{author}{Joost \surnamestart Vennekens\surnameend},
  \bibinfo{author}{Marc \surnamestart Denecker\surnameend} \&
  \bibinfo{author}{Jan \surnamestart {Van den Bussche}\surnameend}
  (\bibinfo{year}{2014}): \emph{\bibinfo{title}{{FO(C)}: A Knowledge
  Representation Language of Causality}}.
\newblock {\slshape \bibinfo{journal}{Theory Pract. Log. Program.}}
  \bibinfo{volume}{14}(\bibinfo{number}{4--5-Online-Supplement}), pp.
  \bibinfo{pages}{60--69}, \doi{10.48550/arXiv.1405.1833}.

\bibitemdeclare{article}{BDKT97AbstractArgumentation-TheoreticApproachDefaultReasoning}
\bibitem{BDKT97AbstractArgumentation-TheoreticApproachDefaultReasoning}
\bibinfo{author}{Andrei \surnamestart Bondarenko\surnameend},
  \bibinfo{author}{Phan~Minh \surnamestart Dung\surnameend},
  \bibinfo{author}{Robert~A. \surnamestart Kowalski\surnameend} \&
  \bibinfo{author}{Francesca \surnamestart Toni\surnameend}
  (\bibinfo{year}{1997}): \emph{\bibinfo{title}{An Abstract,
  Argumentation-Theoretic Approach to Default Reasoning}}.
\newblock {\slshape \bibinfo{journal}{Artif. Intell.}} \bibinfo{volume}{93},
  pp. \bibinfo{pages}{63--101}, \doi{10.1016/S0004-3702(97)00015-5}.

\bibitemdeclare{inproceedings}{BSWW18WeightedAbstractDialecticalFrameworks}
\bibitem{BSWW18WeightedAbstractDialecticalFrameworks}
\bibinfo{author}{Gerhard \surnamestart Brewka\surnameend},
  \bibinfo{author}{Hannes \surnamestart Strass\surnameend},
  \bibinfo{author}{Johannes~Peter \surnamestart Wallner\surnameend} \&
  \bibinfo{author}{Stefan \surnamestart Woltran\surnameend}
  (\bibinfo{year}{2018}): \emph{\bibinfo{title}{Weighted Abstract Dialectical
  Frameworks}}.
\newblock In: {\slshape \bibinfo{booktitle}{Proceedings of the Thirty-Second
  {AAAI} Conference on Artificial Intelligence, ({AAAI}-18), the 30th
  innovative Applications of Artificial Intelligence (IAAI-18), and the 8th
  {AAAI} Symposium on Educational Advances in Artificial Intelligence
  (EAAI-18), New Orleans, Louisiana, {USA}, February 2-7, 2018}}, pp.
  \bibinfo{pages}{1779--1786}, \doi{10.1609/aaai.v32i1.11545}.

\bibitemdeclare{article}{CRS18ApproximationFixpointTheoryWell-FoundedSemanticsHigher-Order}
\bibitem{CRS18ApproximationFixpointTheoryWell-FoundedSemanticsHigher-Order}
\bibinfo{author}{Angelos \surnamestart Charalambidis\surnameend},
  \bibinfo{author}{Panos \surnamestart Rondogiannis\surnameend} \&
  \bibinfo{author}{Ioanna \surnamestart Symeonidou\surnameend}
  (\bibinfo{year}{2018}): \emph{\bibinfo{title}{Approximation Fixpoint Theory
  and the Well-Founded Semantics of Higher-Order Logic Programs}}.
\newblock {\slshape \bibinfo{journal}{Theory Pract. Log. Program.}}
  \bibinfo{volume}{18}(\bibinfo{number}{3-4}), pp. \bibinfo{pages}{421--437},
  \doi{10.1017/S1471068418000108}.

\bibitemdeclare{inproceedings}{DvBJD16CompositionalTypedHigher-OrderLogicDefinitions}
\bibitem{DvBJD16CompositionalTypedHigher-OrderLogicDefinitions}
\bibinfo{author}{Ingmar \surnamestart Dasseville\surnameend},
  \bibinfo{author}{Matthias \surnamestart {van der Hallen}\surnameend},
  \bibinfo{author}{Bart \surnamestart Bogaerts\surnameend},
  \bibinfo{author}{Gerda \surnamestart Janssens\surnameend} \&
  \bibinfo{author}{Marc \surnamestart Denecker\surnameend}
  (\bibinfo{year}{2016}): \emph{\bibinfo{title}{A Compositional Typed
  Higher-Order Logic with Definitions}}.
\newblock In: {\slshape \bibinfo{booktitle}{Technical Communications of the
  32nd International Conference on Logic Programming, {ICLP} 2016 TCs, October
  16-21, 2016, New York City, {USA}}}, pp. \bibinfo{pages}{14:1--14:13},
  \doi{10.4230/OASIcs.ICLP.2016.14}.

\bibitemdeclare{inproceedings}{DBS15FormalTheoryJustifications}
\bibitem{DBS15FormalTheoryJustifications}
\bibinfo{author}{Marc \surnamestart Denecker\surnameend},
  \bibinfo{author}{Gerhard \surnamestart Brewka\surnameend} \&
  \bibinfo{author}{Hannes \surnamestart Strass\surnameend}
  (\bibinfo{year}{2015}): \emph{\bibinfo{title}{A Formal Theory of
  Justifications}}.
\newblock In: {\slshape \bibinfo{booktitle}{Logic Programming and Nonmonotonic
  Reasoning - 13th International Conference, {LPNMR} 2015, Lexington, KY,
  {USA}, September 27-30, 2015. Proceedings}}, pp. \bibinfo{pages}{250--264},
  \doi{10.1007/978-3-319-23264-5\_22}.

\bibitemdeclare{incollection}{DMT00ApproximationsStableOperatorsWell-FoundedFixpointsApplications}
\bibitem{DMT00ApproximationsStableOperatorsWell-FoundedFixpointsApplications}
\bibinfo{author}{Marc \surnamestart Denecker\surnameend},
  \bibinfo{author}{Victor \surnamestart Marek\surnameend} \&
  \bibinfo{author}{Miros{\l}aw \surnamestart Truszczy{\'n}ski\surnameend}
  (\bibinfo{year}{2000}): \emph{\bibinfo{title}{Approximations, Stable
  Operators, Well-Founded Fixpoints and Applications in Nonmonotonic
  Reasoning}}.
\newblock In \bibinfo{editor}{Jack \surnamestart Minker\surnameend}, editor:
  {\slshape \bibinfo{booktitle}{Logic-Based Artificial Intelligence}},
  {\slshape \bibinfo{series}{The Springer International Series in Engineering
  and Computer Science}} \bibinfo{volume}{597}, \bibinfo{publisher}{Springer
  US}, pp. \bibinfo{pages}{127--144}, \doi{10.1007/978-1-4615-1567-8_6}.

\bibitemdeclare{article}{DMT04Ultimateapproximationapplicationnonmonotonicknowledgerepresentation}
\bibitem{DMT04Ultimateapproximationapplicationnonmonotonicknowledgerepresentation}
\bibinfo{author}{Marc \surnamestart Denecker\surnameend},
  \bibinfo{author}{Victor~W. \surnamestart Marek\surnameend} \&
  \bibinfo{author}{Miroslaw \surnamestart Truszczynski\surnameend}
  (\bibinfo{year}{2004}): \emph{\bibinfo{title}{Ultimate approximation and its
  application in nonmonotonic knowledge representation systems}}.
\newblock {\slshape \bibinfo{journal}{Inf. Comput.}}
  \bibinfo{volume}{192}(\bibinfo{number}{1}), pp. \bibinfo{pages}{84--121},
  \doi{10.1016/j.ic.2004.02.004}.

\bibitemdeclare{inproceedings}{DPB01UltimateWell-FoundedStableSemanticsLogicPrograms}
\bibitem{DPB01UltimateWell-FoundedStableSemanticsLogicPrograms}
\bibinfo{author}{Marc \surnamestart Denecker\surnameend},
  \bibinfo{author}{Nikolay \surnamestart Pelov\surnameend} \&
  \bibinfo{author}{Maurice \surnamestart Bruynooghe\surnameend}
  (\bibinfo{year}{2001}): \emph{\bibinfo{title}{Ultimate Well-Founded and
  Stable Semantics for Logic Programs with Aggregates}}.
\newblock In: {\slshape \bibinfo{booktitle}{Logic Programming, 17th
  International Conference, {ICLP} 2001, Paphos, Cyprus, November 26 - December
  1, 2001, Proceedings}}, pp. \bibinfo{pages}{212--226},
  \doi{10.1007/3-540-45635-X\_22}.

\bibitemdeclare{inproceedings}{DV07Well-FoundedSemanticsAlgebraicTheoryNon-monotoneInductive}
\bibitem{DV07Well-FoundedSemanticsAlgebraicTheoryNon-monotoneInductive}
\bibinfo{author}{Marc \surnamestart Denecker\surnameend} \&
  \bibinfo{author}{Joost \surnamestart Vennekens\surnameend}
  (\bibinfo{year}{2007}): \emph{\bibinfo{title}{Well-Founded Semantics and the
  Algebraic Theory of Non-monotone Inductive Definitions}}.
\newblock In: {\slshape \bibinfo{booktitle}{Logic Programming and Nonmonotonic
  Reasoning, 9th International Conference, {LPNMR} 2007, Tempe, AZ, {USA}, May
  15-17, 2007, Proceedings}}, pp. \bibinfo{pages}{84--96},
  \doi{10.1007/978-3-540-72200-7\_9}.

\bibitemdeclare{inproceedings}{DT10TowardsProbabilisticArgumentationJury-basedDisputeResolution}
\bibitem{DT10TowardsProbabilisticArgumentationJury-basedDisputeResolution}
\bibinfo{author}{Phan~Minh \surnamestart Dung\surnameend} \&
  \bibinfo{author}{Phan~Minh \surnamestart Thang\surnameend}
  (\bibinfo{year}{2010}): \emph{\bibinfo{title}{Towards (Probabilistic)
  Argumentation for Jury-based Dispute Resolution}}.
\newblock In: {\slshape \bibinfo{booktitle}{Computational Models of Argument:
  Proceedings of {COMMA} 2010, Desenzano del Garda, Italy, September 8-10,
  2010}}, pp. \bibinfo{pages}{171--182}, \doi{10.3233/978-1-60750-619-5-171}.

\bibitemdeclare{article}{DHMPW11WeightedargumentsystemsBasicdefinitionsalgorithms}
\bibitem{DHMPW11WeightedargumentsystemsBasicdefinitionsalgorithms}
\bibinfo{author}{Paul~E. \surnamestart Dunne\surnameend},
  \bibinfo{author}{Anthony \surnamestart Hunter\surnameend},
  \bibinfo{author}{Peter \surnamestart McBurney\surnameend},
  \bibinfo{author}{Simon \surnamestart Parsons\surnameend} \&
  \bibinfo{author}{Michael~J. \surnamestart Wooldridge\surnameend}
  (\bibinfo{year}{2011}): \emph{\bibinfo{title}{Weighted argument systems:
  Basic definitions, algorithms, and complexity results}}.
\newblock {\slshape \bibinfo{journal}{Artif. Intell.}}
  \bibinfo{volume}{175}(\bibinfo{number}{2}), pp. \bibinfo{pages}{457--486},
  \doi{10.1016/j.artint.2010.09.005}.

\bibitemdeclare{article}{H16AppropriateCausalModelsstabilityCausation}
\bibitem{H16AppropriateCausalModelsstabilityCausation}
\bibinfo{author}{Joseph~Y. \surnamestart Halpern\surnameend}
  (\bibinfo{year}{2016}): \emph{\bibinfo{title}{Appropriate Causal Models and
  the stability of Causation}}.
\newblock {\slshape \bibinfo{journal}{Rev. Symb. Log.}}
  \bibinfo{volume}{9}(\bibinfo{number}{1}), pp. \bibinfo{pages}{76--102},
  \doi{10.1017/S1755020315000246}.

\bibitemdeclare{inproceedings}{LM11SocialAbstractArgumentation}
\bibitem{LM11SocialAbstractArgumentation}
\bibinfo{author}{Jo{\~{a}}o \surnamestart Leite\surnameend} \&
  \bibinfo{author}{Jo{\~{a}}o~G. \surnamestart Martins\surnameend}
  (\bibinfo{year}{2011}): \emph{\bibinfo{title}{Social Abstract
  Argumentation}}.
\newblock In: {\slshape \bibinfo{booktitle}{{IJCAI} 2011, Proceedings of the
  22nd International Joint Conference on Artificial Intelligence, Barcelona,
  Catalonia, Spain, July 16-22, 2011}}, pp. \bibinfo{pages}{2287--2292}.
\newblock \urlprefix\url{http://ijcai.org/Proceedings/11/Papers/381.pdf}.

\bibitemdeclare{inproceedings}{LBCYF16FlexibleApproximatorsApproximatingFixpointTheory}
\bibitem{LBCYF16FlexibleApproximatorsApproximatingFixpointTheory}
\bibinfo{author}{Fangfang \surnamestart Liu\surnameend},
  \bibinfo{author}{Yi~\surnamestart Bi\surnameend},
  \bibinfo{author}{Md.~Solimul \surnamestart Chowdhury\surnameend},
  \bibinfo{author}{Jia{-}Huai \surnamestart You\surnameend} \&
  \bibinfo{author}{Zhiyong \surnamestart Feng\surnameend}
  (\bibinfo{year}{2016}): \emph{\bibinfo{title}{Flexible Approximators for
  Approximating Fixpoint Theory}}.
\newblock In: {\slshape \bibinfo{booktitle}{Advances in Artificial Intelligence
  - 29th Canadian Conference on Artificial Intelligence, Canadian {AI} 2016,
  Victoria, BC, Canada, May 31 - June 3, 2016. Proceedings}}, pp.
  \bibinfo{pages}{224--236}, \doi{10.1007/978-3-319-34111-8\_28}.

\bibitemdeclare{article}{OW11Characterizingstrongequivalenceargumentationframeworks}
\bibitem{OW11Characterizingstrongequivalenceargumentationframeworks}
\bibinfo{author}{Emilia \surnamestart Oikarinen\surnameend} \&
  \bibinfo{author}{Stefan \surnamestart Woltran\surnameend}
  (\bibinfo{year}{2011}): \emph{\bibinfo{title}{Characterizing strong
  equivalence for argumentation frameworks}}.
\newblock {\slshape \bibinfo{journal}{Artif. Intell.}}
  \bibinfo{volume}{175}(\bibinfo{number}{14-15}), pp.
  \bibinfo{pages}{1985--2009}, \doi{10.1016/j.artint.2011.06.003}.

\bibitemdeclare{incollection}{S72Continuouslattices}
\bibitem{S72Continuouslattices}
\bibinfo{author}{Dana \surnamestart Scott\surnameend} (\bibinfo{year}{1972}):
  \emph{\bibinfo{title}{Continuous lattices}}.
\newblock In \bibinfo{editor}{E.~\surnamestart Lawvere\surnameend}, editor:
  {\slshape \bibinfo{booktitle}{Toposes, Algebraic Geometry and Logic}},
  {\slshape \bibinfo{series}{Lecture Notes in Mathematics}}
  \bibinfo{volume}{274}, \bibinfo{publisher}{Springer Verlag}, pp.
  \bibinfo{pages}{97--136}, \doi{10.1007/BFb0073967}.

\bibitemdeclare{article}{S13Approximatingoperatorssemanticsabstractdialecticalframeworks}
\bibitem{S13Approximatingoperatorssemanticsabstractdialecticalframeworks}
\bibinfo{author}{Hannes \surnamestart Strass\surnameend}
  (\bibinfo{year}{2013}): \emph{\bibinfo{title}{Approximating operators and
  semantics for abstract dialectical frameworks}}.
\newblock {\slshape \bibinfo{journal}{Artif. Intell.}} \bibinfo{volume}{205},
  pp. \bibinfo{pages}{39--70}, \doi{10.1016/J.ARTINT.2013.09.004}.

\bibitemdeclare{book}{T91Semanticsprogramminglanguages}
\bibitem{T91Semanticsprogramminglanguages}
\bibinfo{author}{Robert~D. \surnamestart Tennent\surnameend}
  (\bibinfo{year}{1991}): \emph{\bibinfo{title}{Semantics of programming
  languages}}.
\newblock \bibinfo{series}{Prentice Hall International Series in Computer
  Science}, \bibinfo{publisher}{Prentice Hall}.

\bibitemdeclare{article}{DBLP:journals/amai/Truszczynski06}
\bibitem{DBLP:journals/amai/Truszczynski06}
\bibinfo{author}{Miroslaw \surnamestart Truszczynski\surnameend}
  (\bibinfo{year}{2006}): \emph{\bibinfo{title}{Strong and uniform equivalence
  of nonmonotonic theories - an algebraic approach}}.
\newblock {\slshape \bibinfo{journal}{Ann. Math. Artif. Intell.}}
  \bibinfo{volume}{48}(\bibinfo{number}{3-4}), pp. \bibinfo{pages}{245--265},
  \doi{10.1007/S10472-007-9049-2}.

\bibitemdeclare{article}{T06Stronguniformequivalencenonmonotonictheories-}
\bibitem{T06Stronguniformequivalencenonmonotonictheories-}
\bibinfo{author}{Miroslaw \surnamestart Truszczynski\surnameend}
  (\bibinfo{year}{2006}): \emph{\bibinfo{title}{Strong and uniform equivalence
  of nonmonotonic theories - an algebraic approach}}.
\newblock {\slshape \bibinfo{journal}{Ann. Math. Artif. Intell.}}
  \bibinfo{volume}{48}(\bibinfo{number}{3-4}), pp. \bibinfo{pages}{245--265},
  \doi{10.1007/s10472-007-9049-2}.

\bibitemdeclare{inproceedings}{VCBD16DistributedAutoepistemicLogicApplicationAccessControl}
\bibitem{VCBD16DistributedAutoepistemicLogicApplicationAccessControl}
\bibinfo{author}{Pieter \surnamestart {Van Hertum}\surnameend},
  \bibinfo{author}{Marcos \surnamestart Cramer\surnameend},
  \bibinfo{author}{Bart \surnamestart Bogaerts\surnameend} \&
  \bibinfo{author}{Marc \surnamestart Denecker\surnameend}
  (\bibinfo{year}{2016}): \emph{\bibinfo{title}{Distributed Autoepistemic Logic
  and its Application to Access Control}}.
\newblock In: {\slshape \bibinfo{booktitle}{Proceedings of the Twenty-Fifth
  International Joint Conference on Artificial Intelligence, {IJCAI} 2016, New
  York, {NY}, {USA}, 9-15 July 2016}}, pp. \bibinfo{pages}{1286--1292}.
\newblock \urlprefix\url{http://www.ijcai.org/Abstract/16/186}.

\bibitemdeclare{article}{VGD06SplittingoperatorAlgebraicmodularityresultslogics}
\bibitem{VGD06SplittingoperatorAlgebraicmodularityresultslogics}
\bibinfo{author}{Joost \surnamestart Vennekens\surnameend},
  \bibinfo{author}{David \surnamestart Gilis\surnameend} \&
  \bibinfo{author}{Marc \surnamestart Denecker\surnameend}
  (\bibinfo{year}{2006}): \emph{\bibinfo{title}{Splitting an operator:
  Algebraic modularity results for logics with fixpoint semantics}}.
\newblock {\slshape \bibinfo{journal}{{ACM} Trans. Comput. Log.}}
  \bibinfo{volume}{7}(\bibinfo{number}{4}), pp. \bibinfo{pages}{765--797},
  \doi{10.1145/1183278.1183284}.

\bibitemdeclare{article}{VWMD07PredicateIntroductionLogicsFixpointSemanticsI}
\bibitem{VWMD07PredicateIntroductionLogicsFixpointSemanticsI}
\bibinfo{author}{Joost \surnamestart Vennekens\surnameend},
  \bibinfo{author}{Johan \surnamestart Wittocx\surnameend},
  \bibinfo{author}{Maarten \surnamestart Mari{\"{e}}n\surnameend} \&
  \bibinfo{author}{Marc \surnamestart Denecker\surnameend}
  (\bibinfo{year}{2007}): \emph{\bibinfo{title}{Predicate Introduction for
  Logics with a Fixpoint Semantics. Part {I:} Logic Programming}}.
\newblock {\slshape \bibinfo{journal}{Fundam. Informaticae}}
  \bibinfo{volume}{79}(\bibinfo{number}{1-2}), pp. \bibinfo{pages}{187--208},
  \doi{10.3233/FUN-2007-791-209}.


\bibitemdeclare{article}{VWMD07PredicateIntroductionLogicsFixpointSemanticsII}
\bibitem{VWMD07PredicateIntroductionLogicsFixpointSemanticsII}
\bibinfo{author}{Joost \surnamestart Vennekens\surnameend},
  \bibinfo{author}{Johan \surnamestart Wittocx\surnameend},
  \bibinfo{author}{Maarten \surnamestart Mari{\"{e}}n\surnameend} \&
  \bibinfo{author}{Marc \surnamestart Denecker\surnameend}
  (\bibinfo{year}{2007}): \emph{\bibinfo{title}{Predicate Introduction for
  Logics with Fixpoint Semantics. Part {II:} Autoepistemic Logic}}.
\newblock {\slshape \bibinfo{journal}{Fundam. Informaticae}}
  \bibinfo{volume}{79}(\bibinfo{number}{1-2}), pp. \bibinfo{pages}{209--227},
  \doi{10.3233/FUN-2007-791-209}.

\end{thebibliography}

\end{document}